\pdfoutput=1
\documentclass[cernpreprint,texlive=2011,txfonts,UKenglish,PAPER]{atlasdoc}

\usepackage[subfigure]{atlaspackage}
\usepackage{atlasbiblatex}

\usepackage{multirow}
\usepackage [BSM, other,process,particle]{atlasphysics}

\AtlasTitle{Search for Higgs and $Z$ Boson Decays to $\phi\,\gamma$ with the ATLAS Detector}
\author{The ATLAS Collaboration}

\date{\today}
\AtlasRefCode{HIGG-2016-05}

\PreprintIdNumber{CERN-EP-2016-130}

\AtlasJournalRef{Phys. Rev. Lett. 117, 111802}
\AtlasDOI{10.1103/PhysRevLett.117.111802}

\AtlasAbstract{%
A search for the decays of the Higgs and $Z$ bosons to
a $\phi$ meson and a photon is performed with a $pp$
collision data sample corresponding to an integrated luminosity of
$2.7\,\ifb$ collected at $\sqrt{s}=13\TeV$ with the ATLAS detector at the
LHC. No significant excess of events is
observed above the background, and 95\% confidence level upper limits on the
branching fractions of the Higgs and $Z$ boson decays to $\phi\,\gamma$ of $1.4\times10^{-3}$ and
$8.3\times10^{-6}$, respectively, are obtained.
}

\hypersetup{pdftitle={ATLAS document},pdfauthor={The ATLAS Collaboration}}

\newcommand\progname[1]{{\sc #1}}
\def\ee{\ensuremath{e^+ e^-}}%
\def\epem{\ensuremath{e^+ e^-}}%
\def\mumu{\ensuremath{\mathrm{\mu^+ \mu^-}}}%

\def\leplep{\ensuremath{\ell^+ \ell^-}}%

\def\ptmuq#1 {\ensuremath{p_{\textrm{T}}^{\,\mu^{#1}}}}

\def\etamuq#1 {\ensuremath{\eta^{\,\mu^{#1}}}}

\def\ptgamma {\ensuremath{p_{\textrm{T}}^{\gamma}}}
\def\etagamma {\ensuremath{\eta^{\gamma}}}

\def\ptKK{\ensuremath{p_{\mathrm{T}}^{K^{+}K^{-}}}} 
\def\phiKK{\ensuremath{\phi\to K^{+}K^{-}}}
\def\KK{\ensuremath{K^{+}K^{-}}}

\newcommand{\tabscript}[3]{%
  \setlength{\fboxrule}{0pt}%
  \fbox{\ensuremath{#1^{#2}_{#3}}}%

  }
\begin{document}

\maketitle



Rare decays of the $125\;\gev$ Higgs
boson~\cite{ATLAS:2012af,Chatrchyan:2012ia} $H$ to a light meson and a
photon $\gamma$ have been suggested to present one viable probe of the
Yukawa coupling of the Higgs boson to light ($u,d,s$)
quarks~\cite{Kagan:2014ila,Koenig:2015pha,Perez:2015lra}.  While the
Standard Model (SM) predicts these couplings to be small, substantial modifications are
predicted in several scenarios beyond the SM, which include the
Minimal Flavor Violation framework~\cite{D'Ambrosio:2002ex}, the
Froggatt-Nielsen mechanism~\cite{Froggatt:1978nt}, the Higgs-dependent
Yukawa couplings model~\cite{Giudice:2008uua}, the Randall-Sundrum
family of models \cite{Randall:1999ee}, and the possibility of the
Higgs boson being a composite pseudo-Goldstone
boson~\cite{Dugan:1984hq}.
The light-quark Yukawa couplings are almost entirely unconstrained by
existing data and the large multijet background at the Large Hadron
Collider (LHC) severely inhibits the study of such couplings with
inclusive $H\to q\bar{q}$ decays.
The decay of the Higgs boson to a $\phi$ meson and a photon would give
access to the strange-quark Yukawa coupling and to potential
deviations from the SM prediction. The expected SM branching fraction
is ${\cal B}\left(H\to\phi\,\gamma\right)=(2.3\pm0.1)\times
10^{-6}$~\cite{Koenig:2015pha}, and no direct experimental information
about this decay mode currently exists.
The analogous rare decays of the Higgs boson to a heavy quarkonium
state and a photon offer sensitivity to the charm- and bottom-quark
Yukawa couplings~\cite{VatoHBranching,HBranching,Bodwin:2014bpa}. The
Higgs boson decays to $J/\psi\,\gamma$ and $\Upsilon\,\gamma$ have
already been searched for by the ATLAS
collaboration~\cite{Aad:2015sda}. The former decay mode has also been
searched for by the CMS collaboration~\cite{Khachatryan:2015lga}.

The corresponding decay of the $Z$ boson has also been considered from
a theoretical perspective~\cite{Huang:2014cxa,Grossmann:2015lea}, as
it offers a precision test of the SM and the predictions of the
factorization approach in quantum
chromodynamics~\cite{Grossmann:2015lea}. Owing to the large $Z$ boson
production cross section at the LHC, rare $Z$ boson decays can be
probed at branching fractions much smaller than for Higgs boson decays
to the same final state. The most precise prediction for the SM
branching fraction is ${\cal B}\left(Z\to\phi\,\gamma\right)=(1.17\pm0.08)\times
10^{-8}$~\cite{Huang:2014cxa}. The decay $Z\to\phi\,\gamma$ has not
yet been observed and is not well constrained by existing measurements
of $Z$ boson decays.

This Letter describes a search for Higgs and $Z$ boson decays to the
exclusive final state $\phi\,\gamma$. The decay $\phi\to K^{+}K^{-}$
is used to reconstruct the $\phi$ meson. The search is performed with
a sample of $pp$ collision data corresponding to an integrated
luminosity of $2.7\,\ifb$ recorded at a center-of-mass energy
$\rts=13\TeV$ with the ATLAS detector, described in detail in
Ref.~\cite{Aad:2008zzm}.

Higgs~boson production is modeled using the \progname{POWHEG-BOX} v2
Monte Carlo (MC) event
generator~\cite{powheg1,powheg2,powheg3,powheg4,powheg5} for the gluon
fusion ($ggH$) and vector-boson fusion (VBF) processes calculated up
to next-to-leading order in $\alphas$ with CT10 parton distribution
functions~\cite{Lai:2010vv}. Additional contributions from the
associated production of a Higgs boson and a $W$ or $Z$ boson (denoted
$WH$ and $ZH$, respectively) are modeled by the
\progname{Pythia\,8.186} MC event
generator~\cite{Sjostrand:2007gs,Sjostrand:2006za} with NNPDF 2.3
parton distribution functions~\cite{Ball:2012cx}. The production rates
and dynamics for a SM Higgs boson with $\mH = 125\:\gev$, obtained
from Ref.~\cite{Heinemeyer:2013tqa}, are assumed throughout this
analysis. The $ggH$ signal model is appropriately scaled to account
for the production of a Higgs~boson in association with a $t\bar{t}$
or $\bbbar$ pair.
The \progname{POWHEG-BOX} v2 MC event generator, with the CTEQ6L1
parton distribution functions~\cite{Pumplin:2002vw}, is used to model
$Z$ boson production. The total cross section is obtained from the
measurement in Ref.~\cite{Aad:2016naf}, with an uncertainty of 5.5\%.

The Higgs and $Z$ boson decays are simulated as a cascade of two-body
decays. Effects of the helicity of the $\phi$ mesons on the $K^{\pm}$
kinematics are found to modify the acceptance by at most $\pm1\%$ and
this is corrected for in the Higgs boson case and treated as a
systematic uncertainty in the $Z$ boson case, due to the unknown $Z$
boson polarization.

\progname{Pythia\,8.186}~\cite{Sjostrand:2007gs,Sjostrand:2006za} with
the AZNLO set of hadronization and underlying-event
parameters~\cite{Aad:2014xaa} is used to simulate showering and
hadronization.  The simulated events are passed through the detailed
GEANT4 simulation of the ATLAS detector~\cite{GEANT4,2010wqa} and
processed with the same software used to reconstruct data.

The data sample used in this analysis was collected with a dedicated
trigger, commissioned in September 2015, requiring an isolated photon
with a transverse momentum $\pt$ greater than $35\,\GeV$ and an
isolated pair of tracks with an invariant mass loosely consistent with
the $\phi$ meson mass of $1019.5\MeV$~\cite{PDG}, one of which must
have a transverse momentum greater than $15\GeV$. The trigger
efficiency for both the Higgs and $Z$ boson signals is around 80\%
with respect to the offline selection. Events are retained for
analysis if collected under stable LHC beam conditions and the
detector was operating normally.

For this analysis, in the absence of particle identification
capabilities in the relevant momentum range, every reconstructed
charged particle satisfying the following requirements is assumed to
be a $K^{\pm}$ meson.
Events are selected if there are at least two tracks with $\pT > 400
\mev$ originating from the primary vertex, which is defined as the
vertex with the largest $\sum\pT^2$ in the event.  The charged kaons
are reconstructed from inner-detector tracks that satisfy quality
requirements, including a requirement on the number of hits in the
silicon detectors~\cite{ATL-PHYS-PUB-2015-051}.
The $K^{\pm}$ candidates are required to have pseudorapidity
\footnote{ATLAS uses a right-handed coordinate system with its origin at the nominal interaction point (IP) in the center of the detector and the $z$-axis along the beam pipe. The $x$-axis points from the IP to the center of the LHC ring, and the $y$-axis points upward. 
Cylindrical coordinates are used in the transverse plane, with
$\Delta\phi$ being the difference in azimuthal angles around the
$z$-axis.  The pseudorapidity is defined in terms of the polar angle
$\theta$ as $\eta=-\ln\tan(\theta/2)$.}
$|\eta|<2.5$ and $\pt>15\GeV$. The $\phiKK$ decays
are reconstructed from pairs of oppositely charged inner detector
tracks. The higher-$\pt$ track in a pair, denoted the leading track,
is required to have $\pt>20\GeV$. The experimental resolution in
$m_{K^{+}K^{-}}$ is around $4\MeV$, comparable to the natural width of
the $\phi$ meson, $\Gamma_{\phi} = 4.266\pm0.031\MeV$~\cite{PDG}. Track
pairs with a mass $m_{K^{+}K^{-}}$ within $\pm20\MeV$ of the $\phi$
meson mass~\cite{PDG} are selected as $\phiKK$ candidates.
Selected $\phiKK$ candidates are required to satisfy an isolation
requirement: the sum of the $\pt$ of the reconstructed inner detector
tracks from the main vertex within $\Delta R=\sqrt{(\Delta\phi)^{2} +
  (\Delta\eta)^{2}} = 0.2$ of the leading track (excluding both tracks
constituting the $\phiKK$ candidate) is required to be less than 10\%
of the $\pt$ of the $\phi$ candidate, $\ptKK$.

Photons are reconstructed from clusters of energy in the
electromagnetic calorimeter. Clusters without matching tracks are
classified as unconverted photon candidates while clusters matched to
tracks consistent with the hypothesis of a photon conversion into an
$\ee$ pair are classified as converted photon
candidates~\cite{Aaboud:2016yuq}.  Reconstructed photon candidates are
required to have transverse momentum $\ptgamma>35\GeV$, pseudorapidity
$|\etagamma|<2.37$, excluding the barrel/endcap calorimeter transition
region $1.37<|\etagamma|<1.52$, and to satisfy the ``tight'' photon
identification criteria~\cite{Aad:2010sp}. An isolation requirement is
imposed to further suppress the contamination from jets.  The sum of
the transverse momenta of all tracks within $\Delta R = 0.2$ of the
photon direction, excluding those associated with the reconstructed
photon, is required to be less than $5\%$ of $\ptgamma$.  The effects
of multiple $pp$ interactions per bunch crossing~(pile-up) in this
calculation are reduced by removing tracks that do not originate from
the primary vertex.  Additionally, the sum of the transverse momenta
of all energy deposits in the calorimeters within $\Delta R = 0.4$ of
the photon direction, excluding those associated with the
reconstructed photon, is required to be less than $\left(2.45\GeV +
0.022\times\ptgamma\right)$. The calorimeter isolation measurements
are also corrected for the effects of pile-up.

Combinations of a $\phiKK$ candidate and a photon, satisfying
$\Delta\phi(\KK,\gamma) > 0.5$, are retained for further
analysis. When multiple combinations are possible, the combination of
the highest-\pt photon and the $\phiKK$ candidate with a mass closest
to the $\phi$ meson mass is retained. The transverse momentum of
$\phi\to K^{+}K^{-}$ candidates is required to be greater than a
threshold that varies as a function of the invariant mass of the
three-body system, $m_{\KK\gamma}$. Thresholds of $40\GeV$ and
$45\GeV$ are imposed for the regions $m_{\KK\gamma}<91\GeV$ and
$m_{\KK\gamma}\geq125\GeV$, respectively. The threshold is varied from
$40\GeV$ to $45\GeV$ as a linear function of $m_{\KK\gamma}$ in the
region $91\leq m_{\KK\gamma}<125\GeV$. This approach ensures optimal
sensitivity for both the Higgs and $Z$ boson searches.  The total
signal efficiency (kinematic acceptance, and trigger and
reconstruction efficiencies) is 18\% and 8\% for the Higgs and
$Z$~boson decays, respectively. The difference in efficiencies
primarily arises due to the softer $\ptgamma$ and $\ptKK$
distributions in the case of $Z\to\phi\,\gamma$ production. The
$m_{\KK\gamma}$ resolution is around 1.8\% for both the Higgs and
$Z$~boson decays. The $m_{\KK}$ distribution for selected
$\phi\,\gamma$ candidates, with no $m_{\KK}$ requirement applied, is
shown in Figure~\ref{fig:PhiFit} and exhibits a clear peak at the
$\phi$ meson mass.

\begin{figure}[h]
\centering
\includegraphics[width=0.7\linewidth]{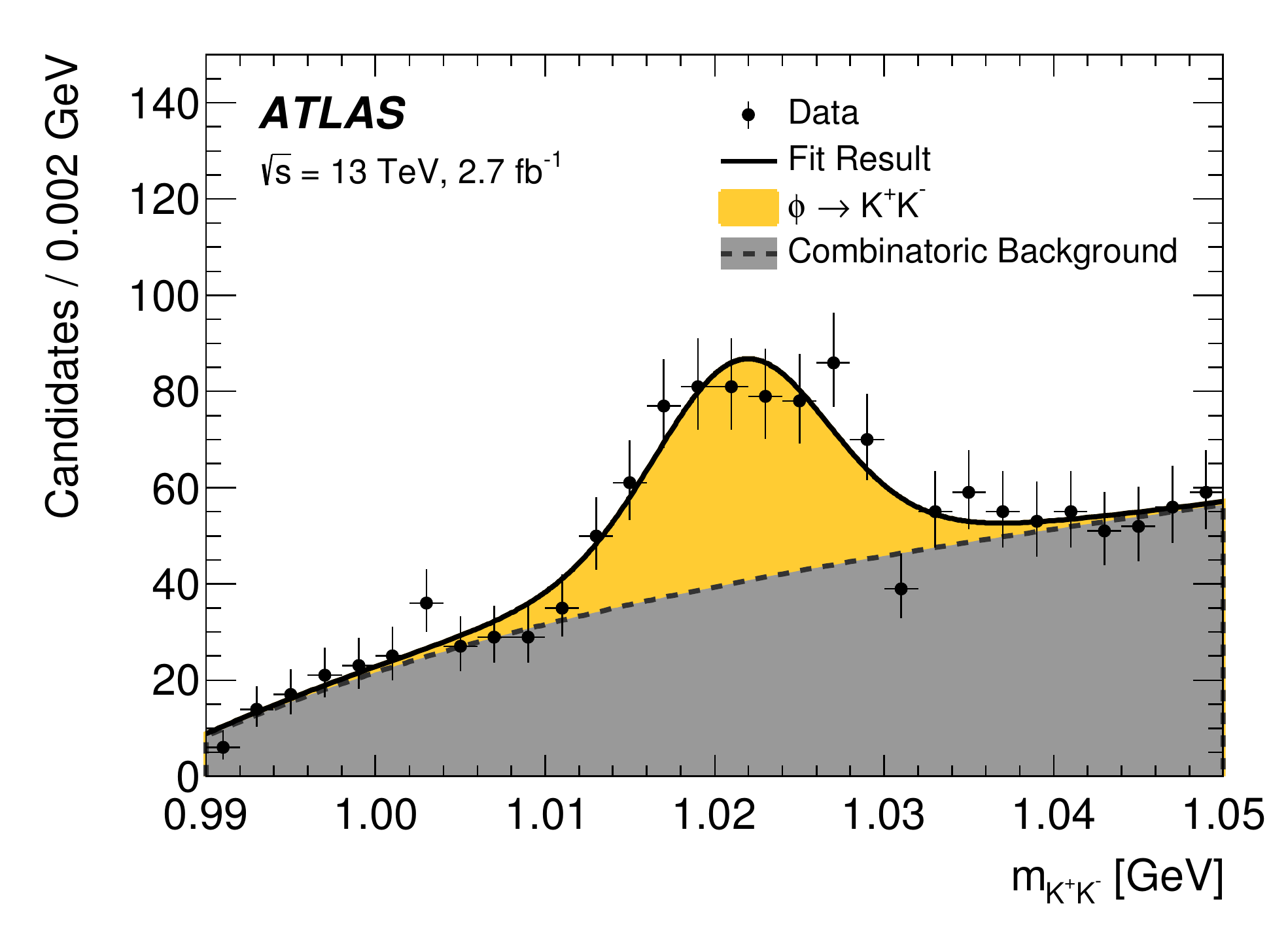}
\caption{The $m_{\KK}$ distribution of selected $\phi\,\gamma$
  combinations with the complete event selection applied (see text),
  apart from the requirement on $m_{\KK}$.  The data are fitted with
  the convolution of a Breit-Wigner distribution, using the $\phi$
  width~\cite{PDG}, and a Gaussian distribution to represent the
  experimental resolution, while the background is modeled with an
  analytical function, commonly used to describe a kinematic
  threshold~\cite{Lees:2013uxa}.
\label{fig:PhiFit}}
\end{figure}

The main source of background to the search comes from events
involving inclusive multijet or photon + jet processes where a
$\phiKK$ candidate is reconstructed from tracks associated with a
jet. The normalization of this inclusive background is extracted
directly from a fit to data. The selection criteria discussed earlier
shape the $m_{\KK\gamma}$ distribution for background such that it
exhibits a threshold structure near $100\GeV$, and falls then smoothly
towards higher mass values.  Given the nontrivial shape of this
background, these processes are modeled with a nonparametric
data-driven approach using templates to describe the kinematic
distributions. A similar procedure was used in the search for Higgs
and $Z$ boson decays to $\Jpsi\,\gamma$ and $\Upsilon(nS)\,\gamma$
described in Ref.~\cite{Aad:2015sda}. The approach exploits a sample
of around 4000 $\KK\gamma$ candidate events passing all of the
kinematic selection requirements described previously, except that the
photon and $\phiKK$ candidates are not required to satisfy the nominal
isolation requirements. The events satisfying this selection are
collected in a generation region (GR). The contamination of this
sample from signal events is expected to be negligible and is verified
not to affect the shape of the background model.
Probability density functions (pdfs) that model the $\ptKK$,
$\ptgamma$, $\Delta\eta(\KK,\gamma)$, and $\Delta\phi(\KK,\gamma)$
distributions of this sample are constructed using a Gaussian kernel
density estimation~\cite{Cranmer:2000du}. Correlations between these
variables and $\ptgamma$ in the event were studied and accounted for
in the background model by deriving separate pdfs in 13 exclusive
regions of $\ptgamma$. In the case of the $\phiKK$ and photon
isolation variables, correlations are accounted for by using
two-dimensional histograms derived in the same 13 exclusive regions of
$\ptgamma$. Values of $m_{\KK}$ are sampled from the corresponding
distribution in the GR. The pdfs of these kinematic and isolation
variables are sampled to generate an ensemble of pseudocandidates,
each with a complete $\KK\,\gamma$ four-vector and an associated pair
of $\phiKK$ and photon isolation values. The nominal selection
requirements are imposed on the ensemble and the surviving
pseudocandidates are used to construct templates for the
$m_{\KK\gamma}$ distribution.

To validate this background model with data, the $m_{\KK\gamma}$
distributions in several validation regions, defined by kinematic and
isolation requirements looser than the nominal signal requirements,
are used to compare the prediction of the background model with the
data.
The $m_{\KK\gamma}$ distribution in one of these validation regions,
defined by the GR selection with the addition of the nominal photon
isolation requirement, is shown in Figure~\ref{fig:BkgdVal_VR2}.
The background model is found to describe the data well, and within
the observed statistical uncertainties. A consistency test of the
background modelling procedure has been performed with a sample of
simulated photon + jet events in place of the data; similarly good
agreement is observed. The robustness of the background model is
further validated by splitting the data into high- and
low-$\pt^{\KK\gamma}$ subsets, that exhibit different threshold
structures, and confirming that the background model describes the
shapes of both $m_{\KK\gamma}$ distributions. Further exclusive
background contributions from $Z\to\ell\ell\gamma$ decays have been
studied but are found to represent a negligible contribution for the
selection requirements and dataset used in this analysis.

\begin{figure}[!htbp]
  \begin{center}
    \includegraphics[width=0.7\textwidth]{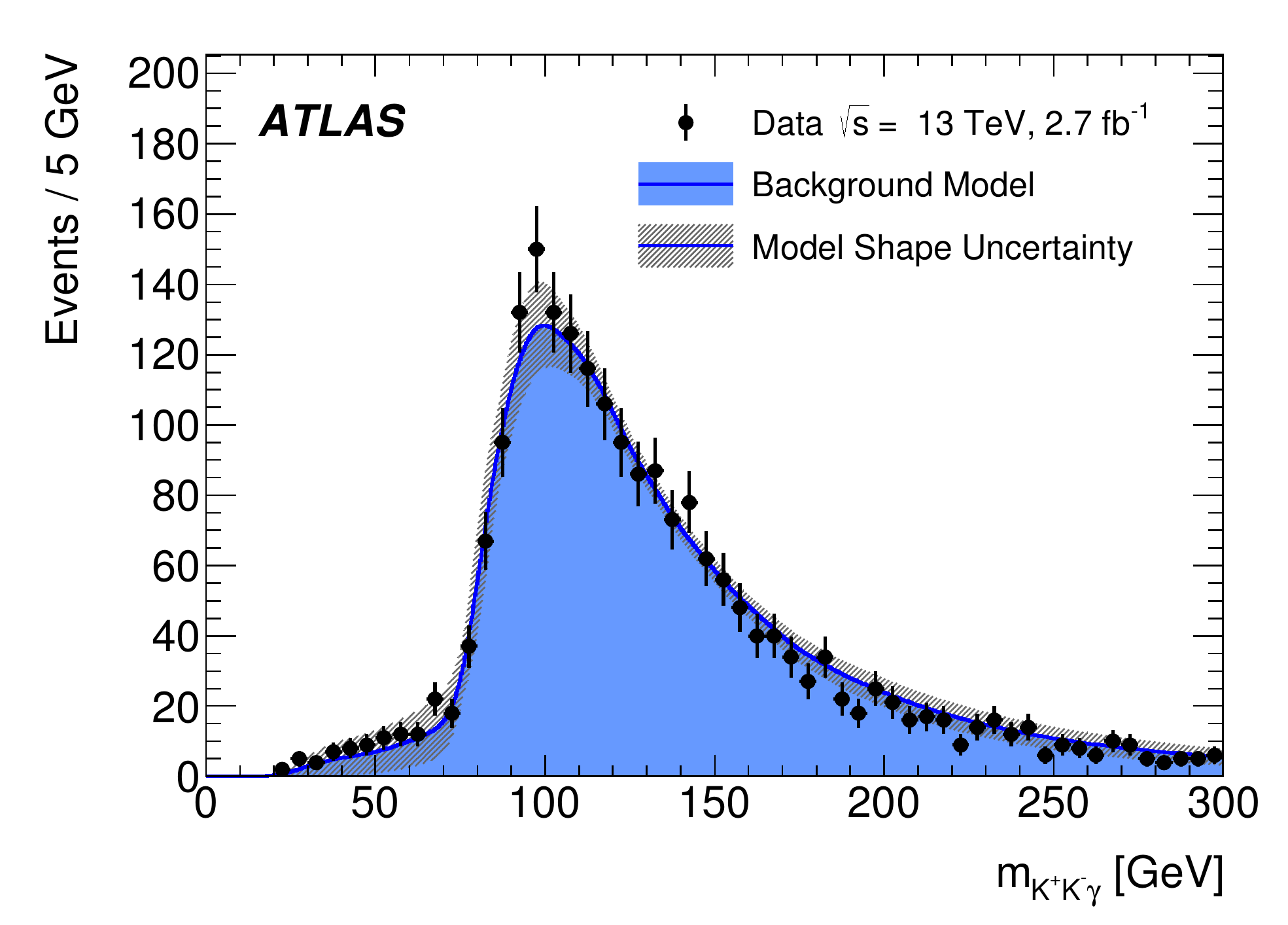}
  \end{center}
  \caption{The distribution of $m_{\KK\gamma}$ in data compared to the
    prediction of the background model for a validation control sample
    defined by the GR selection with the addition of the nominal
    photon isolation requirement. The background model is normalized
    to the observed number of events within the region shown. The
    uncertainty band corresponds to the uncertainty envelope derived
    from variations in the background modeling
    procedure.\label{fig:BkgdVal_VR2} }
\end{figure}

Trigger and identification efficiencies for photons are determined
from samples enriched with $Z\to\ee$ events in
data~\cite{Aaboud:2016yuq,ATLAS:2012cma}. The systematic uncertainty
on the expected signal yield associated with the trigger efficiency is
estimated to be 2\%. The photon identification efficiency
uncertainties, for both the converted and unconverted photons, are
estimated to be 2.4\% and 2.6\% for the Higgs and $Z$~boson signals,
respectively.  An uncertainty of 6\% is assigned to the track
reconstruction efficiency and includes effects associated with the
material budget of the inner detector and the behavior of the track
reconstruction algorithm if a nearby track is present.  The integrated
luminosity of the data sample has an uncertainty of $5\%$ derived
using the method described in Ref.~\cite{Aad:2013ucp}.  The photon
energy scale uncertainty, determined from $Z\to\ee$ events and
validated using $Z\to\ell\ell \gamma$
events~\cite{ref:run1-egamma-calib}, is propagated through the
simulated signal samples as a function of $\etagamma$ and $\ptgamma$.
The uncertainty associated with the description of the photon energy
scale in the simulation is found to be less than 0.3\% of the
three-body invariant mass while the uncertainty associated with the
photon energy resolution is found to be negligible relative to the
overall three-body invariant mass resolution. Similarly, the
systematic uncertainty associated with the track momentum measurement
is found to be negligible.

The uncertainty on the shape of the inclusive multijet and photon +
jet background is estimated through the study of variations in the
background modeling procedure. The shape of the background model is
allowed to vary around the nominal shape within an envelope associated
with shifts in the $\ptKK$ distribution, tilts of the
$\Delta\phi(\KK,\gamma)$ distribution, and by neglecting the weakest
correlation accounted for in the nominal background model.

Results are compared to background and signal predictions using an
unbinned maximum-likelihood fit to the $m_{\KK\gamma}$
distribution. The fit uses the selected events with
$m_{\KK\gamma}<300\:\gev$.
The systematic uncertainties described above result in a 3\%
deterioration of the sensitivity to the $H\to\phi\gamma$ decay.  For
the $Z$ boson decay the reduction is larger, 13\%, mainly due to the
systematic uncertainty in the background shape.
The expected and observed numbers of background events within the
$m_{\KK\gamma}$ ranges relevant to the Higgs and $Z$ boson signals are
shown in Table~\ref{tab:events}.

\begin{table}[h]
\centering
\begin{tabular}{c|cc|cc|c|c}
\hline
\multicolumn{5}{c}{Observed (Expected) Background}& \multicolumn{2}{|c}{Expected Signal} \\ \hline
\multicolumn{5}{c|}{Mass Range [GeV]} & $Z$ & $H$ \\ \cline{1-5}
All & \multicolumn{2}{c|}{$81$--$101$} & \multicolumn{2}{c|}{$120$--$130$}  & ${\cal B} [10^{-6}]$ & ${\cal B} [10^{-3}]$ \\ \hline
1065  & 288 & (266 $\pm$ 9) & 89 & (87 $\pm$ 3) & 6.7 $\pm$ 0.7 & 13.5 $\pm$ 1.5  \\
\hline
\end{tabular}
\caption{The number of observed events and the expected background
  yield for the two $m_{\KK\gamma}$ ranges of interest. The Higgs and
  $Z$~boson contributions expected for branching fraction values of
  $10^{-3}$ and $10^{-6}$, respectively, and estimated using Monte
  Carlo simulations are also shown.\label{tab:events}}
\end{table}

\begin{figure}[h]
\centering
\includegraphics[width=0.7\linewidth]{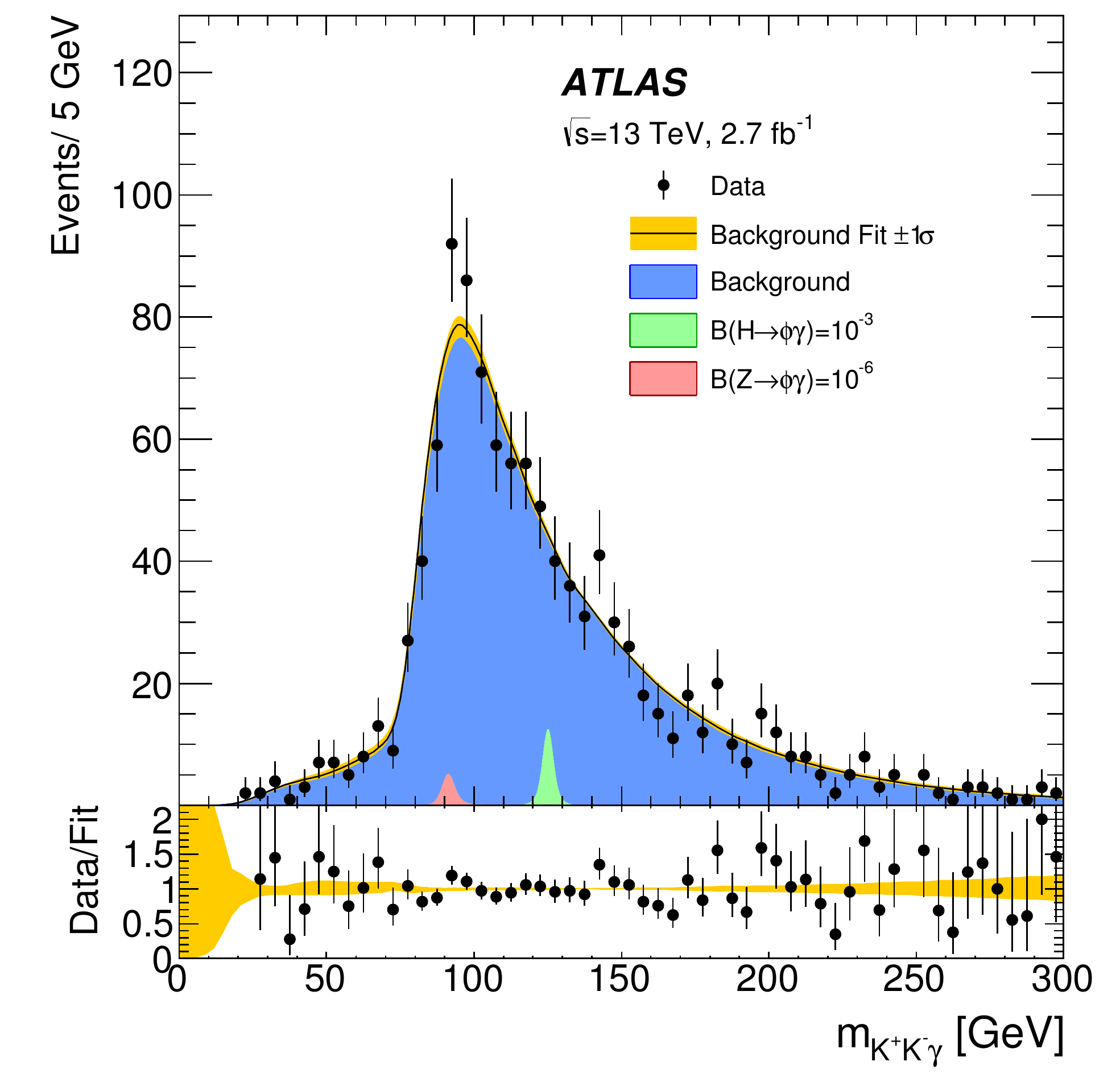}
\caption{The $m_{\KK\gamma}$ distributions of the selected
  $\phi\,\gamma$ candidates, along with the results of the
  maximum-likelihood fit with background-only model. The $1\sigma$
  uncertainty band corresponds to the total uncertainty of the
  background model. The Higgs and $Z$~boson contributions, expected
  for branching fraction values of $10^{-3}$ and $10^{-6}$,
  respectively, are also shown.
\label{fig:Fit}}
\end{figure}

On the basis of the observed data, upper limits are set on the
branching fractions for the Higgs and $Z$~boson decays to
$\phi\,\gamma$ using the CL$_{s}$ modified frequentist
formalism~\cite{cls} with the profile-likelihood ratio test
statistic~\cite{Cowan:2010js}.
The result of the background-only fit is shown in
Figure~\ref{fig:Fit}; a small excess of two standard deviations is
observed in the $Z$~boson mass region, estimated using the asymptotic
approximation for the distribution of the test statistic.
The expected SM production cross section is assumed for the Higgs
boson while the ATLAS measurement of the inclusive $Z$~boson cross
section is used for the $Z$~boson signal~\cite{Aad:2016naf}. The
results are summarized in Table~\ref{fig:Limit}.
The observed 95\% confidence level (CL) upper limits on the branching
fractions for $H\to\phi\,\gamma$ and $Z\to\phi\,\gamma$ decays are
around $600$ and $700$ times the expected SM branching fractions,
respectively.

\begin{table}[h!]
\centering

\begin{tabular}{c|c|c}
\hline
Branching Fraction Limit (95\% CL) & Expected & Observed \\ \hline
${\cal B}\left(H\to\phi\,\gamma\right)[\tabscript{10}{-3}{}]$ & $1.5_{-0.4}^{+0.7}$ & 1.4 \\
${\cal B}\left(Z\to\phi\,\gamma\right)[\tabscript{10}{-6}{}]$ & $4.4_{-1.2}^{+2.0}$ & 8.3 \\
\hline
\end{tabular}
\caption{Expected and observed branching fraction limits at 95\% CL
  for 2.7 fb$^{-1}$ of $pp$ collision data at $\sqrt{s}=13\:\tev$.
  The $\pm 1\sigma$ intervals of the expected limits are also
  given.\label{fig:Limit}}
\end{table}

In conclusion, a search for the decay of Higgs or $Z$ bosons to
$\phi\,\gamma$ has been performed with a $pp$ collision data sample at
$\rts=13\TeV$ corresponding to an integrated luminosity of $2.7\,\ifb$
collected with the ATLAS detector at the LHC. No significant excess of
events is observed above the background. Upper limits at the 95\% CL
are set on the branching fractions for the decay of the $125\;\gev$ SM
Higgs boson and the $Z$ boson to $\phi\,\gamma$. The obtained limits
are ${\cal B}\left(H\to\phi\,\gamma\right)<1.4\times10^{-3}$ and
${\cal B}\left(Z\to\phi\,\gamma\right) < 8.3\times10^{-6}$.

\section*{Acknowledgements}


We thank CERN for the very successful operation of the LHC, as well as the
support staff from our institutions without whom ATLAS could not be
operated efficiently.

We acknowledge the support of ANPCyT, Argentina; YerPhI, Armenia; ARC, Australia; BMWFW and FWF, Austria; ANAS, Azerbaijan; SSTC, Belarus; CNPq and FAPESP, Brazil; NSERC, NRC and CFI, Canada; CERN; CONICYT, Chile; CAS, MOST and NSFC, China; COLCIENCIAS, Colombia; MSMT CR, MPO CR and VSC CR, Czech Republic; DNRF and DNSRC, Denmark; IN2P3-CNRS, CEA-DSM/IRFU, France; GNSF, Georgia; BMBF, HGF, and MPG, Germany; GSRT, Greece; RGC, Hong Kong SAR, China; ISF, I-CORE and Benoziyo Center, Israel; INFN, Italy; MEXT and JSPS, Japan; CNRST, Morocco; FOM and NWO, Netherlands; RCN, Norway; MNiSW and NCN, Poland; FCT, Portugal; MNE/IFA, Romania; MES of Russia and NRC KI, Russian Federation; JINR; MESTD, Serbia; MSSR, Slovakia; ARRS and MIZ\v{S}, Slovenia; DST/NRF, South Africa; MINECO, Spain; SRC and Wallenberg Foundation, Sweden; SERI, SNSF and Cantons of Bern and Geneva, Switzerland; MOST, Taiwan; TAEK, Turkey; STFC, United Kingdom; DOE and NSF, United States of America. In addition, individual groups and members have received support from BCKDF, the Canada Council, CANARIE, CRC, Compute Canada, FQRNT, and the Ontario Innovation Trust, Canada; EPLANET, ERC, FP7, Horizon 2020 and Marie Sk{\l}odowska-Curie Actions, European Union; Investissements d'Avenir Labex and Idex, ANR, R{\'e}gion Auvergne and Fondation Partager le Savoir, France; DFG and AvH Foundation, Germany; Herakleitos, Thales and Aristeia programmes co-financed by EU-ESF and the Greek NSRF; BSF, GIF and Minerva, Israel; BRF, Norway; Generalitat de Catalunya, Generalitat Valenciana, Spain; the Royal Society and Leverhulme Trust, United Kingdom.

The crucial computing support from all WLCG partners is acknowledged gratefully, in particular from CERN, the ATLAS Tier-1 facilities at TRIUMF (Canada), NDGF (Denmark, Norway, Sweden), CC-IN2P3 (France), KIT/GridKA (Germany), INFN-CNAF (Italy), NL-T1 (Netherlands), PIC (Spain), ASGC (Taiwan), RAL (UK) and BNL (USA), the Tier-2 facilities worldwide and large non-WLCG resource providers. Major contributors of computing resources are listed in Ref.~\cite{ATL-GEN-PUB-2016-002}.

\printbibliography



\newpage
\begin{flushleft}
{\Large The ATLAS Collaboration}

\bigskip

M.~Aaboud$^\textrm{\scriptsize 135d}$,
G.~Aad$^\textrm{\scriptsize 86}$,
B.~Abbott$^\textrm{\scriptsize 113}$,
J.~Abdallah$^\textrm{\scriptsize 64}$,
O.~Abdinov$^\textrm{\scriptsize 12}$,
B.~Abeloos$^\textrm{\scriptsize 117}$,
R.~Aben$^\textrm{\scriptsize 107}$,
O.S.~AbouZeid$^\textrm{\scriptsize 137}$,
N.L.~Abraham$^\textrm{\scriptsize 149}$,
H.~Abramowicz$^\textrm{\scriptsize 153}$
% ...
\bigskip
\\
% ...
$^{12}$ Institute of Physics, Azerbaijan Academy of Sciences, Baku, Azerbaijan\\
% ...
$^{64}$ University of Iowa, Iowa City IA, United States of America\\
% ...
$^{86}$ CPPM, Aix-Marseille Universit{\'e} and CNRS/IN2P3, Marseille, France\\
% ...
$^{107}$ Nikhef National Institute for Subatomic Physics and University of Amsterdam, Amsterdam, Netherlands\\
% ...
$^{113}$ Homer L. Dodge Department of Physics and Astronomy, University of Oklahoma, Norman OK, United States of America\\
% ...
$^{117}$ LAL, Univ. Paris-Sud, CNRS/IN2P3, Universit{\'e} Paris-Saclay, Orsay, France\\
% ...
$^{135}$ $^{(a)}$ Facult{\'e} des Sciences Ain Chock, R{\'e}seau Universitaire de Physique des Hautes Energies - Universit{\'e} Hassan II, Casablanca; $^{(b)}$ Centre National de l'Energie des Sciences Techniques Nucleaires, Rabat; $^{(c)}$ Facult{\'e} des Sciences Semlalia, Universit{\'e} Cadi Ayyad, LPHEA-Marrakech; $^{(d)}$ Facult{\'e} des Sciences, Universit{\'e} Mohamed Premier and LPTPM, Oujda; $^{(e)}$ Facult{\'e} des sciences, Universit{\'e} Mohammed V, Rabat, Morocco\\
% ...
$^{137}$ Santa Cruz Institute for Particle Physics, University of California Santa Cruz, Santa Cruz CA, United States of America\\
% ...
$^{149}$ Department of Physics and Astronomy, University of Sussex, Brighton, United Kingdom\\
% ...
$^{153}$ Raymond and Beverly Sackler School of Physics and Astronomy, Tel Aviv University, Tel Aviv, Israel\\
% ...
$^{a}$ Also at Department of Physics, King's College London, London, United Kingdom\\
$^{b}$ Also at Institute of Physics, Azerbaijan Academy of Sciences, Baku, Azerbaijan\\
$^{c}$ Also at Novosibirsk State University, Novosibirsk, Russia\\
$^{d}$ Also at TRIUMF, Vancouver BC, Canada\\
$^{e}$ Also at Department of Physics {\&} Astronomy, University of Louisville, Louisville, KY, United States of America\\
% ...
\end{flushleft}


\end{document}